\documentclass[article]{./macros/elsarticle_qh}

\usepackage[hmarginratio=1:1,top=32mm,left=20mm,columnsep=1cm]{geometry} %
\usepackage{bm}
\usepackage{amsmath}
\usepackage{relsize} 
\usepackage[svgnames]{xcolor} 
\usepackage{grffile} 
\usepackage{graphicx}

\usepackage{hyperref}
\hypersetup{
	colorlinks=true,                          
	linkcolor=DarkRed,
	citecolor=DarkRed,
	urlcolor=DarkRed  }

\usepackage{bm}
\usepackage{amsmath}
\usepackage{booktabs}
\usepackage{amssymb}
\usepackage{etoolbox}
\usepackage{mathtools}
\usepackage{gensymb}
\usepackage[section]{placeins} 
\usepackage{float}
\usepackage{subfig}
\usepackage[thinc]{esdiff}

\DeclareMathAlphabet{\mathsfbi}{OT1}{\sfdefault}{bx}{sl}
\DeclareMathVersion{sfletters}
\SetSymbolFont{letters}{sfletters}{OML}{ntxsfmi}{b}{it}

\makeatletter
\newcommand{\mathbfsbilow}[1]{%
	\text{\mathversion{sfletters}$\m@th#1$}%
}
\DeclareRobustCommand{\tensor}[1]{%
	\begingroup
	\ifcat\noexpand #1\relax
	\edef\greek@test{\detokenize{#1}}%
	\edef\greek@test{\expandafter\@cdr\greek@test\@nil}%
	\edef\greek@test{\expandafter\@car\greek@test\@nil}%
	\edef\x{\the\lccode\expandafter`\greek@test}%
	\edef\y{\number\expandafter`\greek@test}%
	\ifnum\x=\y\relax
	\mathbfsbilow{#1}%
	\else
	\mathsfbi{#1}%
	\fi
	\else
	\mathsfbi{#1}%
	\fi
	\endgroup
}
\makeatother
\makeatletter
\newcommand{\sbullet}{%
	\hbox{\fontfamily{lmr}\fontsize{.4\dimexpr(\f@size pt)}{0}\selectfont\textbullet}}

\makeatother

\usepackage[numbers]{natbib}

\begin{document}
	
\begin{frontmatter}
		
\title{{\Large \textbf{Dispersion relation and wave equation for attenuative elliptically anisotropic media}}
}

\cortext[mycorrespondingauthor]{Corresponding author}

\address[JLU]{College of Geoexploration Science and Technology, Jilin University, Changchun, 130026, P. R. China}
\address[CSM]{Department of Geophysics, Colorado School of Mines, Golden, 80401, USA}

\author[JLU]{Qi Hao\corref{mycorrespondingauthor}} 
\ead{{xqi.hao@gmail.com}}

\author[CSM]{Ilya Tsvankin}
\ead{ilya@mines.edu}

\begin{abstract}
The P-wave slowness and group-velocity surfaces in elliptically anisotropic media are ellipsoids. Elliptical anisotropy is convenient to use as the reference medium in perturbation methods designed to solve P-wave wave-propagation problems for transverse isotropy (TI).  Here, we make the elliptically anisotropic TI model attenuative and discuss the corresponding P-wave dispersion relation and the wave equation. Our analysis leads to two conditions in terms of the Thomsen-type parameters, which guarantee that the P-wave slowness surface and the dispersion relation satisfy elliptical equations. We also obtain the viscoacoustic wave equation for such elliptically anisotropic media and solve it for point-source radiation using the correspondence principle. Numerical examples validate the proposed elliptical conditions and illustrate the behavior of the P-wavefield in attenuative elliptical TI models. 
\end{abstract}


\end{frontmatter}


\section*{Introduction}
Transversely isotropic media are widely used to represent sedimentary formations in seismic inversion and processing.  TI models can be  conveniently described by the well-known Thomsen parameters \cite[]{thomsen:1986,tsvankin:2012}.  For general transverse isotropy, only the SH-wave slowness satisfies an elliptical equation, which also makes the SH wavefront elliptical.  A special type of TI media is elliptical anisotropy, which requires that the Thomsen parameters satisfy the condition $\epsilon=\delta$ \cite[e.g.,][]{tsvankin:2012}. In that case, the P-wave slowness and group-velocity surfaces become ellipsoids, whereas the SV-wave slowness surface is spherical.  

Compared with isotropy, elliptical anisotropy has an additional degree of freedom to account for the difference between the P-wave horizontal and vertical velocities (assuming the symmetry axis to be vertical). Elliptical models can be perturbed to efficiently solve modeling and inverse problems for general TI media \cite[e.g.,][]{slawinski:2004,danek:2012}. 
For example, elliptical anisotropy can be employed as the reference medium to obtain analytic expressions for P-wave traveltimes in general TI models \cite[e.g.,][]{stovas:2012}. Published research on elliptical anisotropy includes analysis of traveltimes \cite[]{helbig:1983,rogister:2005}, reflection/transmission coefficients  \cite[]{levin:1978,daley:1979}, and development of P-wave imaging algorithms \cite[]{verwest:1989,schleicher:2008}. 

Intrinsic attenuation of seismic waves caused by energy absorption in subsurface formations is ubiquitous in the Earth. In particular, attenuative TI models have been used to process seismic data from shales \cite[e.g.,][]{zhu.tsvankin:2007b,zhu:2007,shekar:2011,shekar:2012}. Such signatures as the ray (group) velocity, attenuation coefficient, and geometrical spreading can be obtained from the Christoffel (plane-wave) equation and the point-source solution of the wave equation for atenuative TI media \cite[]{vavryvcuk:2007asympt,shekar.tsvankin:2014,hao.alkhalifah:2019}. The kinematics and geometric spreading of P- and S-waves in attenuative anisotropic media are weakly dependent on the quality factor, unless  attenuation is uncommonly strong \cite[]{vavrycuk:2007,behura:2009b}. The normalized phase and group attenuation coefficients in dissipative TI and orthorhombic media can be conveniently defined in terms of the Thomsen-type attenuation parameters \cite[]{zhu.tsvankin:2006,zhu.tsvankin:2007,behura:2009b}.  

Wave propagation in attenuative elliptically anisotropic media has not attracted much attenuation in the literature. \cite{hao.tsvankin:2022} find that the P-wave anisotropy in viscoacoustic constant-$Q$  VTI (transversely isotropic with a vertical symmetry axis) media becomes elliptical under two conditions expressed in terms of the Thomsen-type parameters. However, those conditions have not been verified for dissipative models that do not have a constant-$Q$ stiffness matrix. This paper analyses the dispersion relation and formulates the elliptical conditions for general attenuative TI media. Also, we present the viscoacoustic wave equation and derive its point-source solution for such elliptical TI models. 

First, we introduce the complete set of the Thomsen and Thomsen-type parameters for an attenuative VTI medium. Then, we derive the conditions for elliptical anisotropy by analyzing the corresponding dispersion relation. The  viscoacoustic wave equation and its point-source solution for elliptically anisotropic media are discussed next. This is followed by numerical examples designed to validate the elliptical conditions and analyze P-wave propagation in constant-$Q$ elliptical VTI models. 

\section{Thomsen-type notation for attenuative transverse isotropy}
The complex stiffness matrix $ \mathbf{M}$ for viscoelastic VTI media is given by \cite[e.g.,][]{zhu.tsvankin:2006}:
\begin{equation} \label{eq:Mgen}
\small{
\mathbf{M} =
\left(
\begin{matrix}
M_{11} & M_{11}-2M_{66} & M_{13} & 0 & 0 & 0 \\
M_{11}-2M_{66} & M_{11} & M_{13} & 0 & 0 & 0 \\
M_{13} & M_{13} & M_{33} & 0 & 0 & 0 \\
0              & 0              & 0              & M_{55} & 0 & 0 \\
0              & 0              & 0              & 0 & M_{55} & 0 \\
0              & 0              & 0              & 0 & 0 & M_{66}
\end{matrix}
\right),
}
\end{equation}
where the stiffness coefficients $M_{ij}$ are expressed through the quality-factor-matrix elements $Q_{ij} \equiv M_{ij}^R / M_{ij}^I$ as:
\begin{equation}
M_{ij} = M_{ij}^R \left[ 1 - i \, \text{sgn}(f)/Q_{ij}  \right] ;
\end{equation}
$f$ is the frequency and $M_{ij}^R$ and $M_{ij}^I$ are the real and imaginary parts of $M_{ij}$, respectively. The minus sign in front of $\text{sgn}(f)$ follows from the Fourier-transform convention in \cite{cerveny:2001}. 

An attenuative VTI medium can be described using Thomsen-type notation \cite[]{zhu.tsvankin:2006,tsvankin.grechka:2011}.  The \cite{thomsen:1986} velocity parameters  \citep[see][]{tsvankin:2012} are defined in the nonattenuative reference VTI medium described by the real stiffness coefficients $M_{ij}^R$. The Thomsen-type attenuation parameters \cite[]{zhu.tsvankin:2006} conveniently characterize the normalized phase attenuation coefficient $\mathcal{A} \equiv |\mathbf{k}_{I}|/|\mathbf{k}_{R}|$ for P-, SV-, and SH-waves, where $\mathbf{k}_{R}$ and $\mathbf{k}_{I}$ are the real and imaginary parts of the complex wave vector. Below we assume that $\mathbf{k}_{R}$ is parallel to $\mathbf{k}_{I}$ (i.e., the inhomogeneity angle is equal to zero). The word ``normalized'' is omitted below for brevity. 

The definitions of the Thomsen and Thomsen-type parameters for P- and SV-waves are listed below. 

The parameter $V_{P0}$ is the vertical velocity of P-waves: 
\begin{equation} \label{eq:vp0def}
V_{P0}\equiv \sqrt{\frac{M_{33}^{R}}{\rho}} ,
\end{equation} 
where $\rho$ denotes density. 

The parameter $V_{S0}$ is the vertical velocity of S-waves:
\begin{equation} \label{eq:vs0def}
V_{S0} \equiv \sqrt{\frac{M_{55}^{R}}{\rho}} .
\end{equation} 

The parameter $\epsilon$ is approximately equal to the fractional difference between the horizontal and vertical P-wave velocities:
\begin{equation} \label{eq:epsdef}
\epsilon \equiv \frac{M_{11}^{R} - M_{33}^{R}}{2M_{33}^{R}} .
\end{equation}

The parameter $\delta$ determines the second derivative of the P-wave phase velocity at vertical incidence: 
\begin{equation} \label{eq:deltadef} 
\def \tmpa { \left( M_{13}^{R} +M_{55}^{R}  \right)^2 }
\def \tmpb { \left( M_{33}^{R} -M_{55}^{R}  \right)^2  }
\def \tmpc { 2M_{33}^{R} (M_{33}^{R} - M_{55}^{R}) }
\delta  \equiv \frac{\tmpa - \tmpb}{\tmpc}  .
\let \tmpa \undefined
\let \tmpb \undefined
\let \tmpc \undefined
\end{equation}

The parameter $\mathcal{A}_{P0}$ is the vertical attenuation coefficient of P-waves:
\begin{equation} \label{eq:Ap0def}
\mathcal{A}_{P0} \equiv Q_{33} \left( \sqrt{1 + \frac{ 1 }{Q_{33}^2}} - 1 \right) \approx 
\frac{ 1 }{2Q_{33}} .
\end{equation} 

The parameter $\mathcal{A}_{S0}$ is the vertical attenuation coefficient of S-waves: 
\begin{equation} \label{eq:As0def}
\mathcal{A}_{S0} \equiv Q_{55} \left( \sqrt{1 + \frac{ 1 }{Q_{55}^2}} - 1 \right) \approx
\frac{ 1 }{2Q_{55}} .
\end{equation} 

The parameter  $\epsilon_{Q}$ is close to the fractional difference between the P-wave horizontal and vertical attenuation coefficients and can be expressed through $Q_{ij}$ as: 
\begin{equation} \label{eq:epsQdef}
\epsilon_{Q} \equiv \frac{Q_{33} - Q_{11}}{Q_{11}} .
\end{equation}

The parameter $\delta_{Q}$ controls the second derivative of the P-wave  phase attenuation coefficient $\mathcal{A}_{P}$ at vertical incidence (note that the first derivative of $\mathcal{A}_{P}$ goes to zero): 
\begin{equation}   \label{eq:deltaQdef}
\def \tmpa { \frac{Q_{33}-Q_{55}}{Q_{55}} }
\def \tmpb { M_{55}^{R} \frac{\left( M_{13}^{R}
+ M_{33}^{R} \right)^{2}}{M_{33}^{R}-M_{55}^{R}} }
\def \tmpc { 2\frac{Q_{33}-Q_{13}}{Q_{13}} }
\def \tmpd { M_{13}^{R} (M_{13}^{R} + M_{55}^{R}) }
\def \tmpe { M_{33}^{R} (M_{33}^{R} - M_{55}^{R}) }
\delta_{Q}  \equiv \frac{\tmpa \tmpb + \tmpc \tmpd}{\tmpe}  .
\let \tmpa \undefined
\let \tmpb \undefined
\let \tmpc \undefined
\let \tmpd \undefined
\let \tmpe \undefined
\end{equation}

\section{Dispersion relation and elliptical conditions}
According to \cite{zhu.tsvankin:2006}, the complex Christoffel equation for plane P- and SV-waves propagating in the $[x,z]$-plane of attenuative VTI media is given by:
\begin{equation} \label{eq:christ}
\left[ M_{11} k_{x}^2 + M_{55} k_{z}^2 - \rho \, \omega^2 \right] \left[ 
M_{55} k_{x}^2 + M_{33} k_{z}^2 - \rho \, \omega^2 \right]  \\
- \left( M_{13} + M_{55} \right)^2 k_{x}^2 \, k_{z}^2 = 0, 
\end{equation}
where $\omega$ is the angular frequency and $k_{x}$ and $k_{z}$ are the $x$- and $z$-components of the complex wave vector $\mathbf{k}$.  

The P-wave dispersion relation can be obtained from equation \ref{eq:christ}:
\begin{equation} \label{eq:wk_vti} 
\def \tmpa  { (M_{11} + M_{55}) k_{x}^2 + (M_{33} + M_{55}) k_{z}^2  } 
\def \tmpb  { (M_{11} - M_{55}) k_{x}^2  + (M_{33} - M_{55}) k_{z}^2   }
\begin{aligned}
\rho \, \omega^2 &= \frac{1}{2} \tmpa \\
                     &+ \frac{1}{2}  \sqrt{\left[ \tmpb \right]^2 + 4\Lambda  k_{x}^2  \, k_{z}^2 } \, ,
\end{aligned}
\let \tmpa \undefined 
\end{equation}
where
\begin{equation} \label{eq:Lam}
\Lambda = (M_{13} + M_{55})^2 - (M_{11} - M_{55}) (M_{33} - M_{55}) .
\end{equation}

By setting $\Lambda = 0$, the dispersion relation can be reduced to an elliptical equation for the complex wavenumber:
\begin{equation} \label{eq:wk_ellip}
\rho \, \omega^2 = M_{11} \, k_{x}^2 + M_{33} \, k_{z}^2   . 
\end{equation}

Using the parameter definitions from the previous section, condition \ref{eq:Lam} can be rewritten as:
\begin{equation} \label{eq:lambda}
\Lambda = \Lambda_{0} + i \Lambda_{1} Q_{33}^{-1} - \Lambda_{2} \, Q_{33}^{-2} , 
\end{equation}
where
\begin{align}
& \Lambda_{0} = 2 \rho^2 V_{P0}^4 \, c \, (\epsilon - \delta) ,  \\
& \Lambda_{1} = \rho^2 V_{P0}^4 \left[-2b (\epsilon - \delta) - c e \right], 
\end{align}
and
\begin{equation}
\Lambda_{2} = \frac{\rho^2 V_{P0}^4}{c (c + 2\delta) }
 \left\{ -a \, c \, d (c + 2\delta) + b^2 \delta^2 + b \, c \, \delta \, \delta_{Q} + c^2 \left[ \frac{1}{4}\delta_{Q}^2 + d (2\delta + \delta_{Q} ) \right] \right\} .  
\end{equation}
The coefficients $a$, $b$, $c$, $d$, and $e$ are given by:
\begin{align}
& a = \epsilon_{Q} + 2\epsilon (1 + \epsilon_{Q}) ,  \\
& b = c + d , \\
& c = 1 - g , \\
& d = 1 - g g_{Q} , \\
& e = \epsilon_{Q}(1+2\epsilon) - \delta_{Q} , 
\end{align}
where
\begin{align}
\label{eq:g}
&g = \frac{V_{S0}^2}{V_{P0}^2} , \\
\label{eq:gQ}
&g_{Q} =\frac{Q_{33}}{Q_{55}} . 
\end{align}
For typical large values of $Q$ ($Q_{33} \gg 1$), the term proportional to $1/Q_{33}^2$ on the right hand side of equation \ref{eq:lambda} can be ignored.  This means that $\Lambda$ vanishes when $\Lambda_{0} = \Lambda_{1} = 0$, which requires that
 \begin{align}
\label{eq:ellip_cond1}
& \epsilon = \delta , \\
\label{eq:ellip_cond2}
& \epsilon_{Q} = \frac{\delta_{Q}}{1 + 2\delta} \, . 
\end{align}
For weak velocity anisotropy ($|\epsilon|,|\delta| \ll 1$), equation \ref{eq:ellip_cond2} reduces to
 \begin{equation} \label{eq:ellip_cond3}
\epsilon_{Q} = \delta_{Q} \, . 
\end{equation}
Equation \ref{eq:ellip_cond1} is the well-known elliptical condition for the slowness surface \cite[e.g.,][]{tsvankin:2012}. It remains valid for attenuative media if the influence of absorption on velocity is weak.  

To understand the meaning of equation \ref{eq:ellip_cond3}, we examine the linearized P-wave phase attenuation coefficient $\mathcal{A}_{P}$  \cite[]{zhu.tsvankin:2006}: 
\begin{equation} \label{eq:Ap} 
\mathcal{A}_{P} = \mathcal{A}_{P0}  (1 + \delta_{Q} \sin^2 \theta \cos^2 \theta + \epsilon_{Q} \sin^4 \theta ) ,  
\end{equation}
where $\theta$ is the phase angle measured from the symmetry axis. Equation \ref{eq:Ap} is obtained under the assumption of weak attenuation and weak velocity and attenuation anisotropy. 

Substituting condition \ref{eq:ellip_cond3} into equation \ref{eq:Ap} and squaring the result yields: 
\begin{equation} \label{eq:Ap2} 
\mathcal{A}_{P}^2 = \mathcal{A}_{P0}^2  (1 +2\epsilon_{Q} \sin^2 \theta ) ,  
\end{equation}
where the term proportional to $\epsilon_{Q}^2$ is dropped according to the assumption of small attenuation.
Comparison with the well known P-wave phase-velocity equation for elliptical media indicates that the inverse of the coefficient $\mathcal{A}_{P}$ described by equation \ref{eq:Ap2} satisfies an elliptical equation.  

Equations \ref{eq:ellip_cond1} and \ref{eq:ellip_cond2} ensure that the dispersion relation is described by the elliptical equation \ref{eq:wk_ellip}.  For viscoacoustic constant-$Q$ media, if equations \ref{eq:ellip_cond1} and \ref{eq:ellip_cond2} are satisfied at the reference frequency, they remain valid at all frequencies \cite[]{hao.tsvankin:2022}. 

\section{Wave equation and its point-source solution}
Using the dispersion relation in equation \ref{eq:wk_ellip} yields the following viscoacoustic wave equation for the wavefield $P$ in elliptically anisotropic media: 
\begin{equation} \label{eq:weq}
\frac{\partial^{2} P}{ \partial t^{2}} =
   \phi_{11} \odot \left( \frac{\partial^{2} P}{\partial x^{2}}  +  \frac{\partial^{2} P}{\partial y^{2}}  \right)
  + \phi_{33} \odot \frac{\partial^{2} P}{\partial z^{2}}  +  S(t) \delta(\mathbf{x}) ,
\end{equation}
where $\delta(\mathbf{x})$ is the 3D Dirac delta function \cite[]{arfken:2013}, $\phi_{ii}$ are the density-normalized relaxation functions, and  $S(t)$ is the source wavelet.  The source is located at the coordinate origin. The operator $\odot$ is the Riemann-Stieltjes convolution integral \cite[]{hudson:1980},  which is described in \cite{hao.greenhalgh:2019,hao.greenhalgh:2021b,hao.greenhalgh:2021a}. 

The frequency-domain version of equation \ref{eq:weq} can be written as:
\begin{equation} \label{eq:weq_freq}
-\omega^2 \hat{P} = m_{11}  \left( \frac{\partial^2 \hat{P}}{\partial x^2} + \frac{\partial^2 \hat{P}}{\partial y^2} \right) +  m_{33} \frac{\partial^2 \hat{P}}{\partial z^2}  + \hat{S}(\omega)  \delta(\mathbf{x}) , 
\end{equation}
where $\hat{P}$ is the frequency-domain wavefield (the spectrum of $P$), $\hat{S}(\omega)$ is the spectrum of the source wavelet, and $m_{ii}$ are the density-normalized complex stiffness coefficients (see equation \ref{eq:Mgen}). The relationship between $m_{ii}$ and $\phi_{ii}$ can be found in \cite{hao.greenhalgh:2021b}. 

Using the correspondence principle \cite[]{ben-menahem.singh:1981,carcione:2014} and the solution of the isotropic acoustic wave equation \cite[]{aki.richards:2002}, we obtain the following solution of the wave equation \ref{eq:weq_freq} (see Appendix A):  
\begin{equation} \label{eq:P}
\hat{P} = \frac{\hat{S}(\omega)}{4\pi  m_{11} \sqrt{m_{33}} \, \tau}  \exp(i \omega \tau ) , 
\end{equation}
where $\tau$ is the complex traveltime from the source to the receiver,
\begin{equation} \label{eq:tau}
\tau = \sqrt{\frac{x^2 + y^2}{m_{11}} + \frac{z^2}{m_{33}} } \, .
\end{equation}
The time-domain solution for point-source radiation is given by the inverse Fourier transform of equation \ref{eq:P}, with the sign convention from \cite{cerveny:2001} and \cite{hao.greenhalgh:2021b,hao.greenhalgh:2021a}. 

The complex traveltime can be expressed in terms of the propagation distance, group velocity, and group attenuation coefficient. As mentioned before, for media with weak attenuation and weak attenuation anisotropy, the influence of attenuation on phase and group velocity can be ignored \cite[e.g.,][]{zhu.tsvankin:2006}. The group attenuation coefficient is close to the phase attenuation coefficient at zero inhomogeneity angle, as proved by \cite{behura:2009b}. Then, the point-source wavefield \ref{eq:P} can be approximately represented as:
\begin{equation} \label{eq:Pnew}
\hat{P} = \frac{\hat{S}(\omega)}{4\pi \, G \, R}  \exp \left(-|\omega|  \mathcal{A}_{P} \, \frac{R}{V_{P}} \right)  \exp\left[i\omega  \frac{R}{V_{P}}  + i \, \text{sgn}(\omega) \beta \right] ,
\end{equation}
where $R$ is the source-receiver distance. The group velocity $V_{P}$ in elliptical media is given by \cite[e.g.,][]{tsvankin:2012}: 
\begin{equation} \label{eq:Vp_ellip_exact}
V_{P} = V_{P0} \left( \cos^2 \psi +  \frac{\sin^2 \psi}{1 + 2\epsilon}  \right)^{-\frac{1}{2}} ,
\end{equation}
where $\psi$ is the group angle measured from the symmetry axis.  

In the linear approximation with respect to $\epsilon$, equation \ref{eq:Vp_ellip_exact} reduces to:
\begin{equation} \label{eq:Vp_ellip}
V_{P} = V_{P0} \, (1 + \epsilon \sin^2\psi) . 
\end{equation}
Equation \ref{eq:Vp_ellip_exact} describes the exact group velocity in the nonattenuative reference medium, whereas equation \ref{eq:Vp_ellip} is identical to the linearized P-wave phase velocity under the elliptical condition $\epsilon=\delta$ \cite[equation \ref{eq:ellip_cond1};][]{tsvankin:2012}. 

The quantity $\mathcal{A}_{P}$ is the group attenuation coefficient, 
\begin{equation} \label{eq:Ap_ellip}
\mathcal{A}_{P} = \mathcal{A}_{P0} (1 + \epsilon_{Q} \sin^2\psi) ,
\end{equation}
which coincides with the linearized phase attenuation coefficient $\mathcal{A}_{P}$  (see equation \ref{eq:Ap}) under the approximate elliptical condition \ref{eq:ellip_cond3}. 

The quantity $G$ is proportional to the relative geometric spreading \cite[]{cerveny:2001,stovas:2018}:
\begin{equation} \label{eq:Gp}
G = V_{P0}^2  \left[ 1 + \epsilon \left( 1 + \cos^2\psi \right) \right] , 
\end{equation}
whereas $\beta$ is the phase shift due to attenuation: 
\begin{equation} \label{eq:beta}
\beta = \mathcal{A}_{P0} \left[ 2 + \epsilon_{Q} \left( 1 + \cos^2\psi \right) \right] .
\end{equation}
If $A_{P0}$ is set to zero, equation \ref{eq:Pnew} reduces to the point-source solution of the wave equation for P-waves in nonattenuative elliptically anisotropic media \cite[]{tsvankin:1995,tsvankin:2012}. 

\section{Numerical examples}
First, we validate the elliptical conditions for attenuative VTI media by examining the phase velocity $V_{P}$ and the phase attenuation coefficient $\mathcal{A}_{P}$ computed from the complex wavenumber \cite[]{zhu.tsvankin:2006,vavrycuk:2007}. In the first scenario (A), we use the elliptical dispersion relation \ref{eq:wk_ellip} to compute the complex P-wavenumber, and then the corresponding  values of $V_{P}$ and $\mathcal{A}_{P}$ from the exact equations. For the second scenario (B), the elliptical conditions \ref{eq:ellip_cond1} and \ref{eq:ellip_cond2} are applied to the general VTI dispersion relation \ref{eq:wk_vti}, with subsequent calculation of the corresponding $V_{P}$ and $\mathcal{A}_{P}$. The third scenario (C) is similar to B, but the dispersion relation \ref{eq:wk_vti} is computed with the conditions \ref{eq:ellip_cond1} and \ref{eq:ellip_cond3}. In the last two scenarios (B and C), the parameter $\delta$ is set equal to $\epsilon$, and $\delta_{Q}$ is calculated from equations \ref{eq:ellip_cond2} and \ref{eq:ellip_cond3}, respectively. 

\begin{figure}[H]
\centering
\subfloat[]{\includegraphics[width=2.1in]{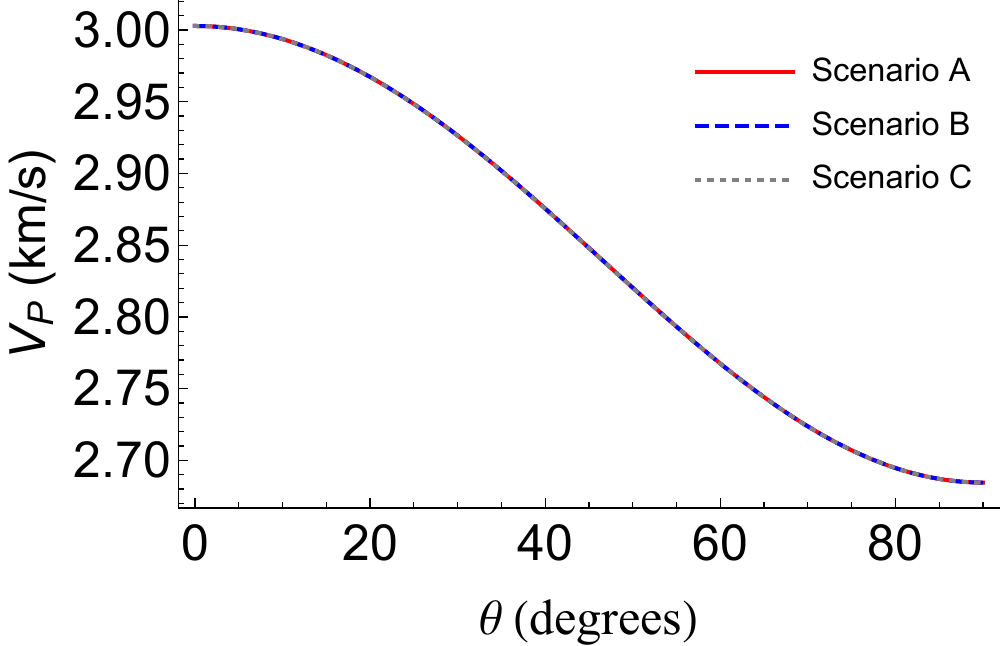}} \qquad
\subfloat[]{\includegraphics[width=2.1in]{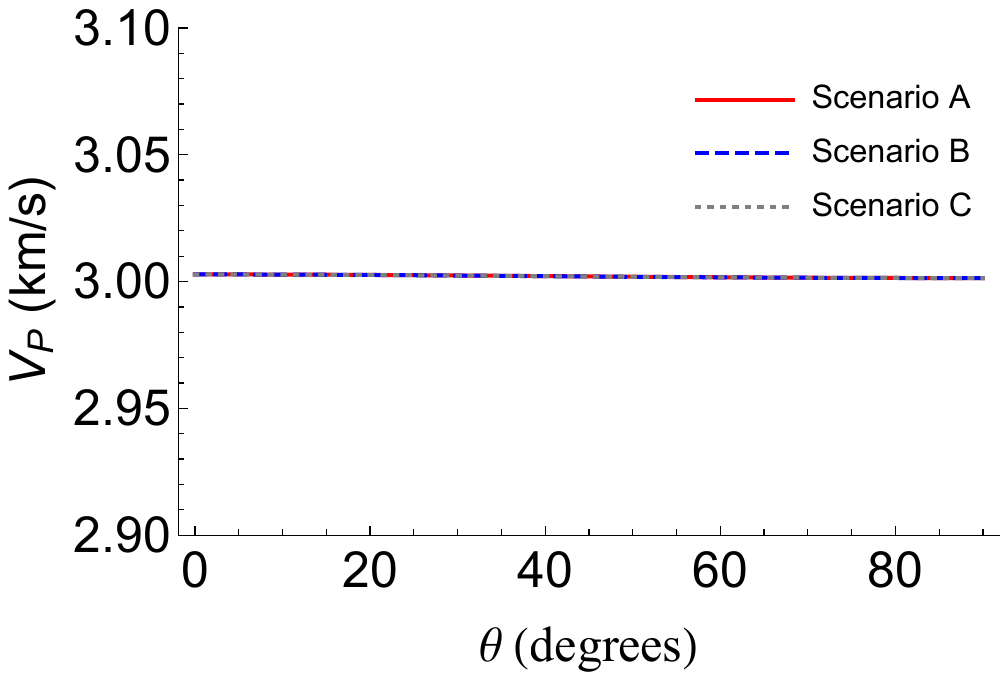}}  \\
\subfloat[]{\includegraphics[width=2.1in]{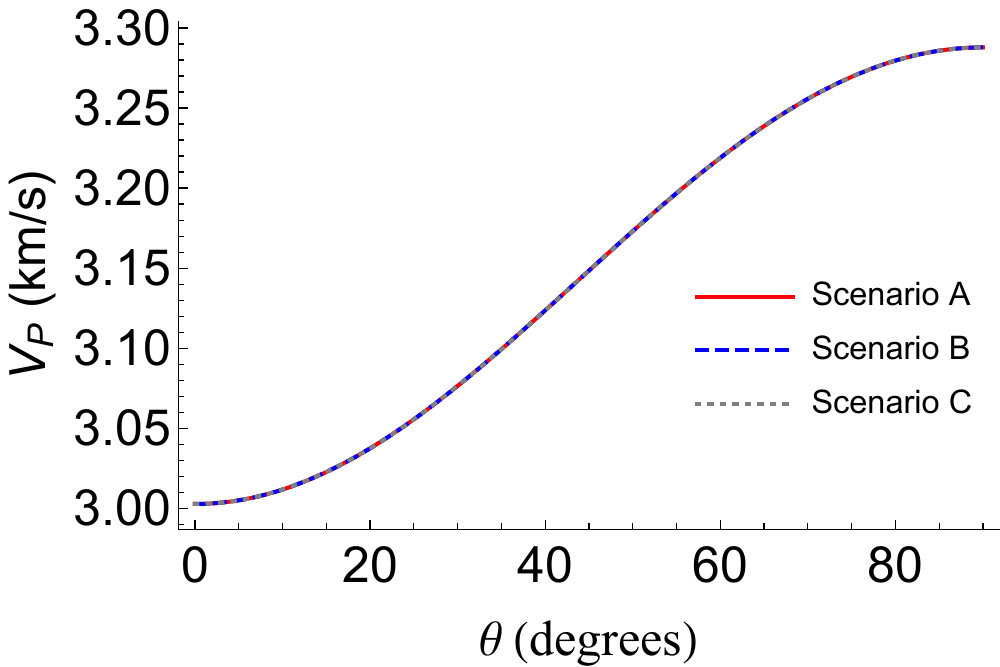}} \qquad
\subfloat[]{\includegraphics[width=2.1in]{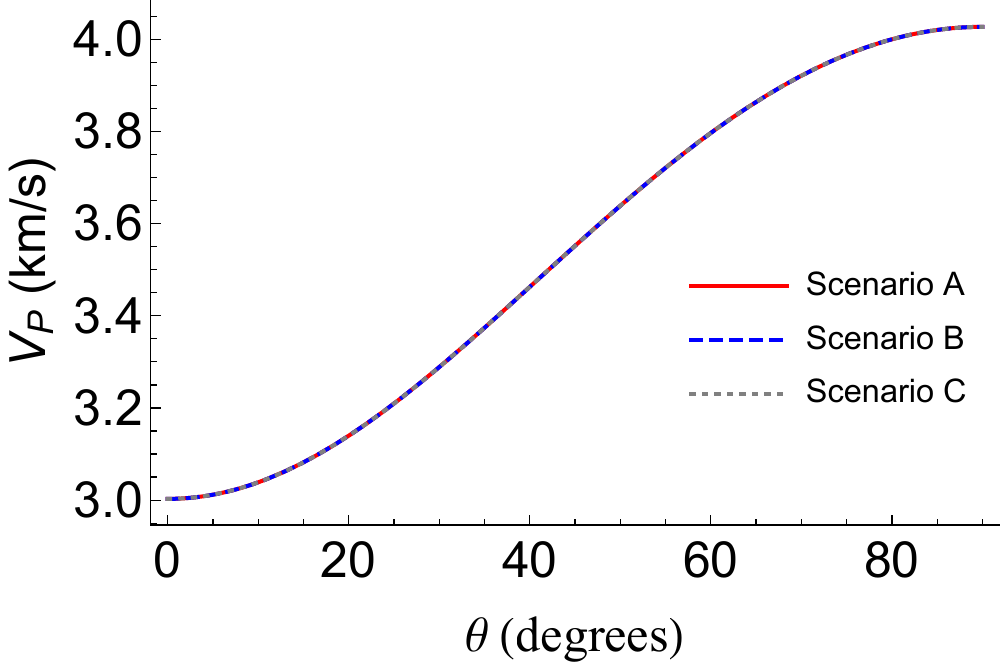}} 
\caption{
Exact P-wave phase velocity for the three scenarios of elliptical conditions described in the text. The medium parameters are $V_{P0}=3$~km/s, $g=0.4$ (see equation \ref{eq:g}), $A_{P0}=0.025$ (corresponding to $Q_{33}=20$), $g_{Q}=4$ (see equation \ref{eq:gQ}), and $\epsilon_{Q}=-0.3$. The parameter $\epsilon$ is  (a) $-0.1$, (b) $0$, (c) $0.1$, and (d) $0.4$. 
}
\label{fig:Vp}
\end{figure}

The phase velocities for all three  scenarios practically coincide (Figures \ref{fig:Vp}).  The normalized attenuation coefficients $\mathcal{A}_{P}$ for scenarios A and B are virtually identical, whereas $\mathcal{A}_{P}$ for scenario C is slightly different, and that deviation increases with $|\epsilon| = |\delta|$. These results confirm that the elliptical conditions \ref{eq:ellip_cond1} and \ref{eq:ellip_cond2} lead to the elliptical dispersion relation. The approximate condition \ref{eq:ellip_cond3} is sufficiently accurate for typical values of the anisotropy coefficients, although it produces a small error that increases with the magnitude of velocity anisotropy.  

\begin{figure}[H]
\centering
\subfloat[]{\includegraphics[width=2.1in]{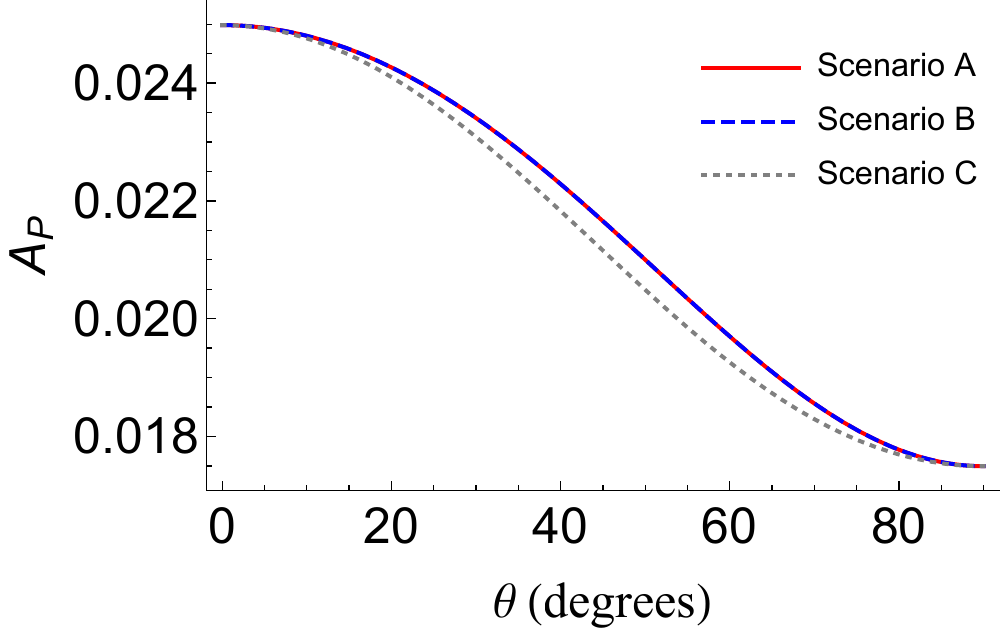}} \qquad
\subfloat[]{\includegraphics[width=2.1in]{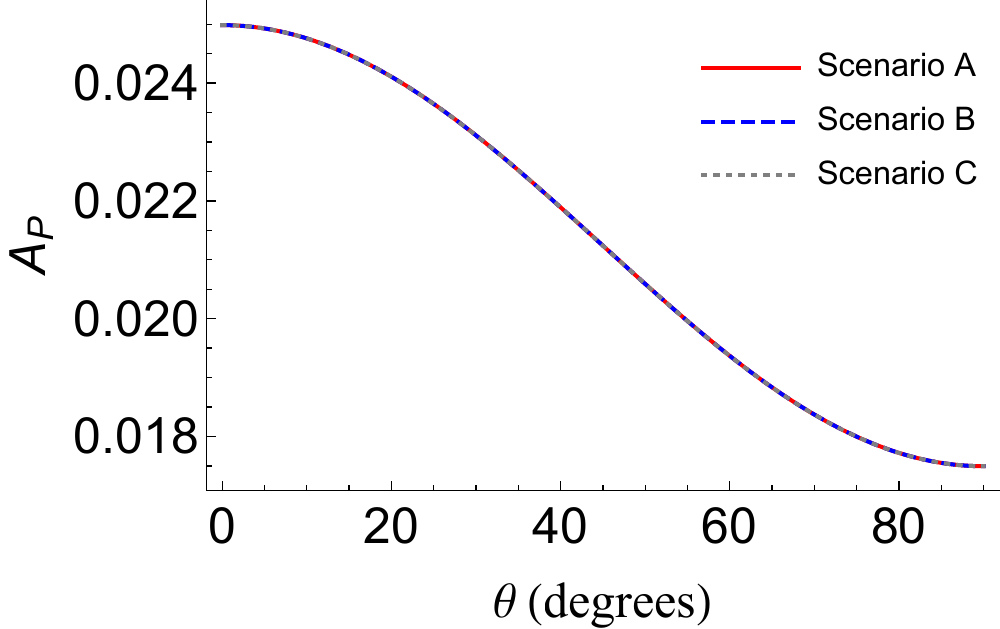}}  \\
\subfloat[]{\includegraphics[width=2.1in]{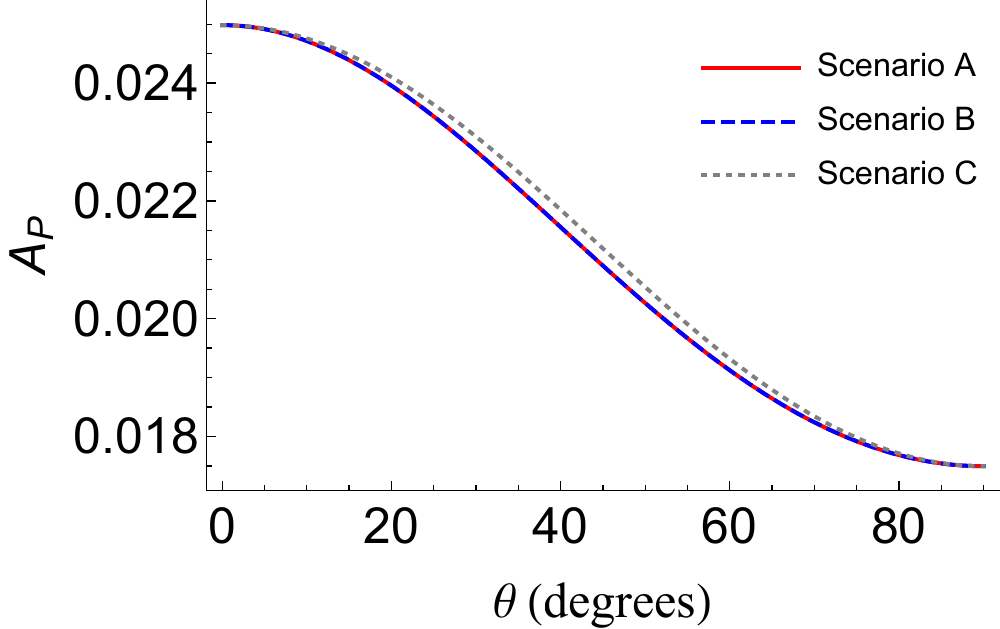}} \qquad
\subfloat[]{\includegraphics[width=2.1in]{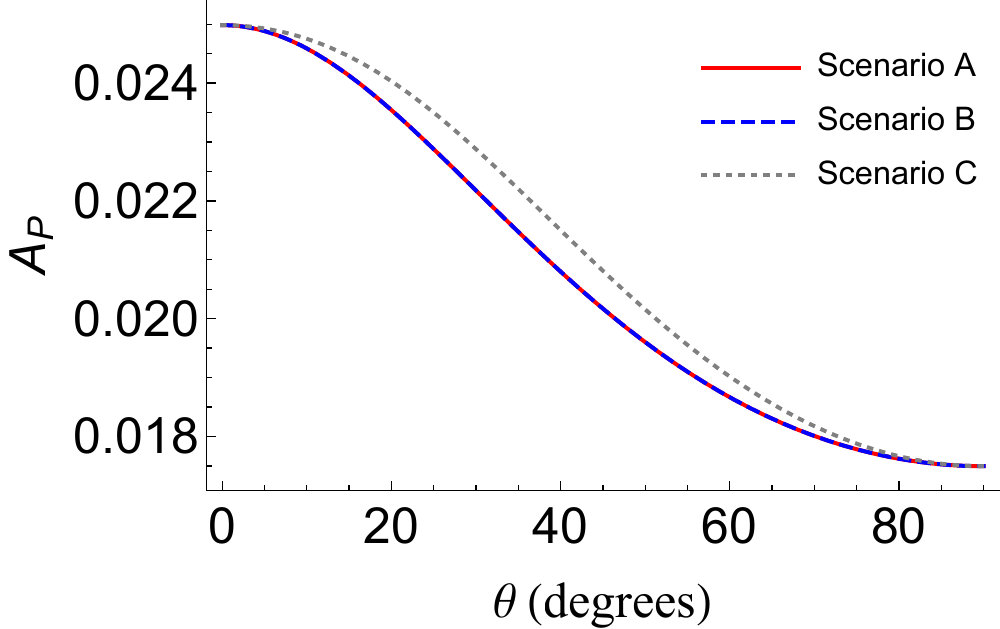}} 
\caption{
Exact P-wave phase attenuation coefficient for the three scenarios of elliptical conditions.  The medium parameters are the same as in Figure \ref{fig:Vp}. 
}
\label{fig:Ap}
\end{figure}

Next, we analyze P-wave waveforms from a point source in a constant-$Q$ elliptically anisotropic VTI medium constructed using Kjartansson's \cite{kjartansson:1979} dissipative model \citep[see][]{hao.tsvankin:2022}. The complex stiffness coefficients $M_{11}$ and $M_{33}$ for that model are given by:
\begin{equation}  \label{eq:Mkjar}
M_{ii} = \frac{\tilde{M}^{R}_{ii}}{\text{cos}(\pi \gamma_{ii}) }
 \left(-i\frac{f}{f_{0}} \right)^{2\gamma_{ii}} ,
\end{equation}
with
\begin{equation} \label{eq:gam}
	\gamma_{ii} = \frac{1}{\pi} \, \text{tan}^{-1} \left( \frac{1}{Q_{ii}} \right) ,
\end{equation}
where $f$ is the frequency, $f_{0}$ is the reference frequency, and $\tilde{M}^{R}_{ii}$ are the real parts of the stiffness coefficients $M_{ii}$ at $f=f_{0}$ ($\tilde{M}^{R}_{ii}=\text{Re}(M_{ii})|_{f=f_0}$).  

\begin{figure}[H]
\centering
\subfloat[]{\includegraphics[width=2.1in]{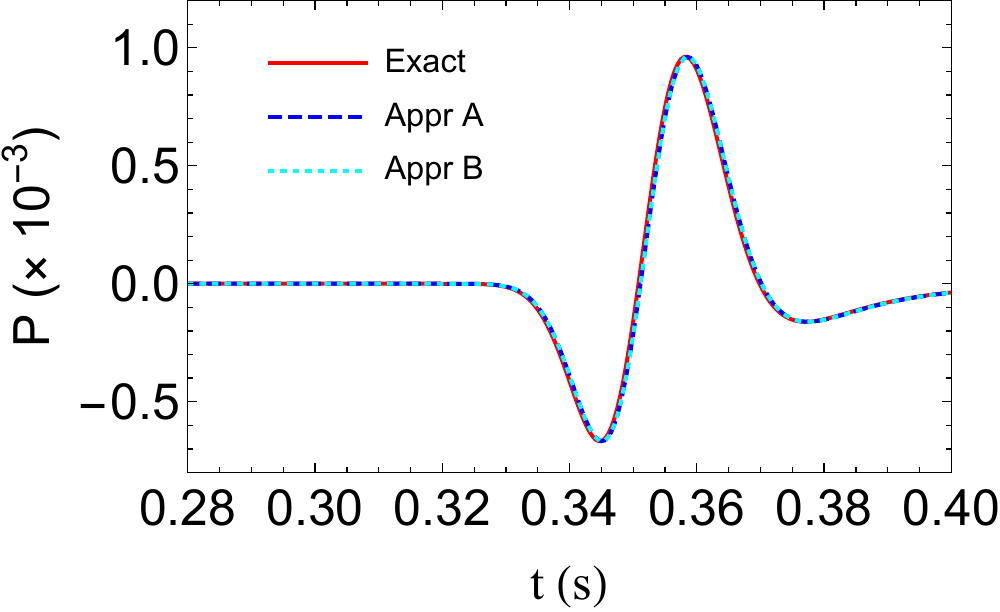}}  \qquad
\subfloat[]{\includegraphics[width=2.1in]{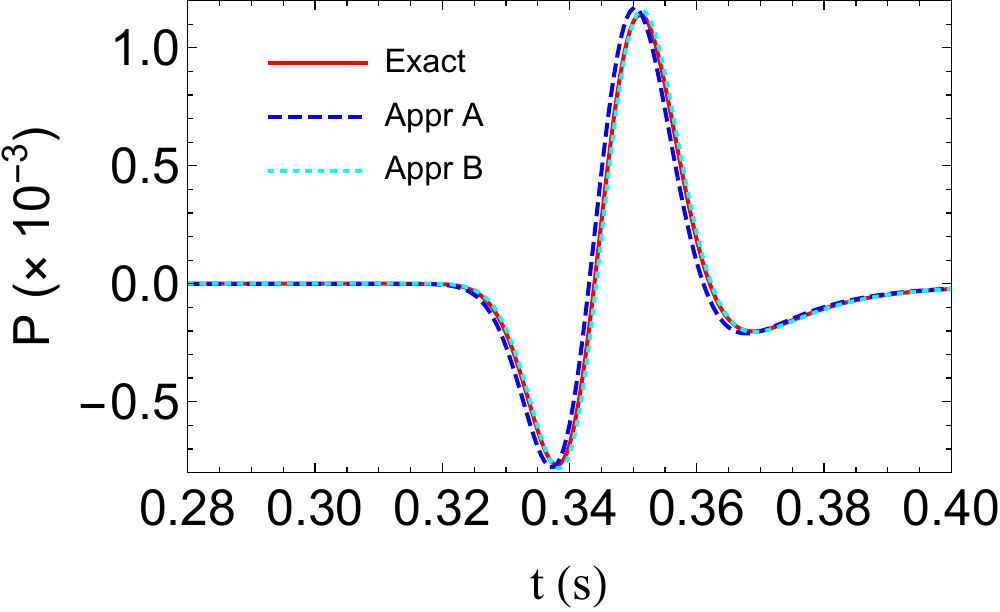}}  \\
\subfloat[]{\includegraphics[width=2.1in]{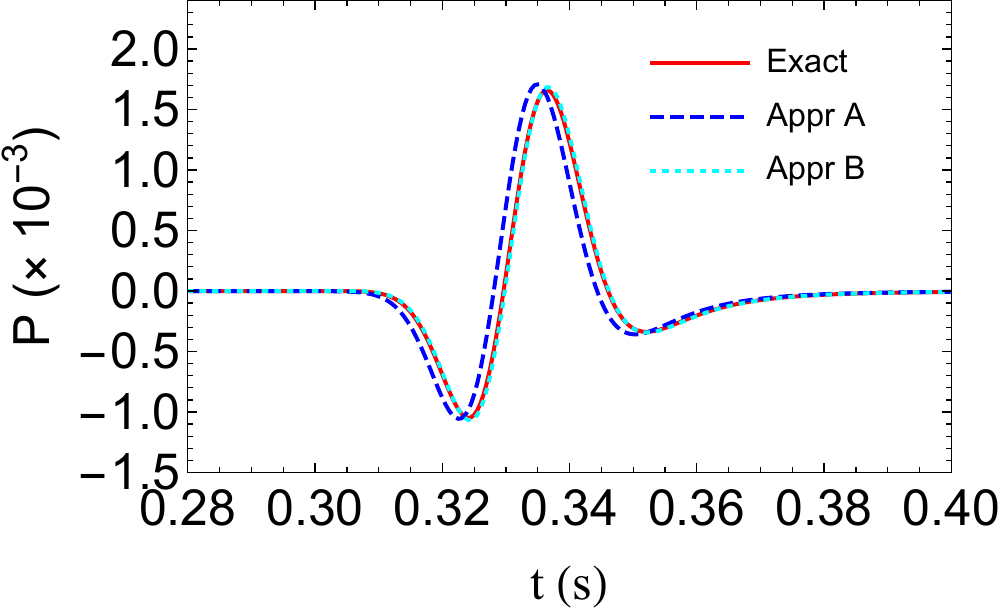}}  \qquad
\subfloat[]{\includegraphics[width=2.1in]{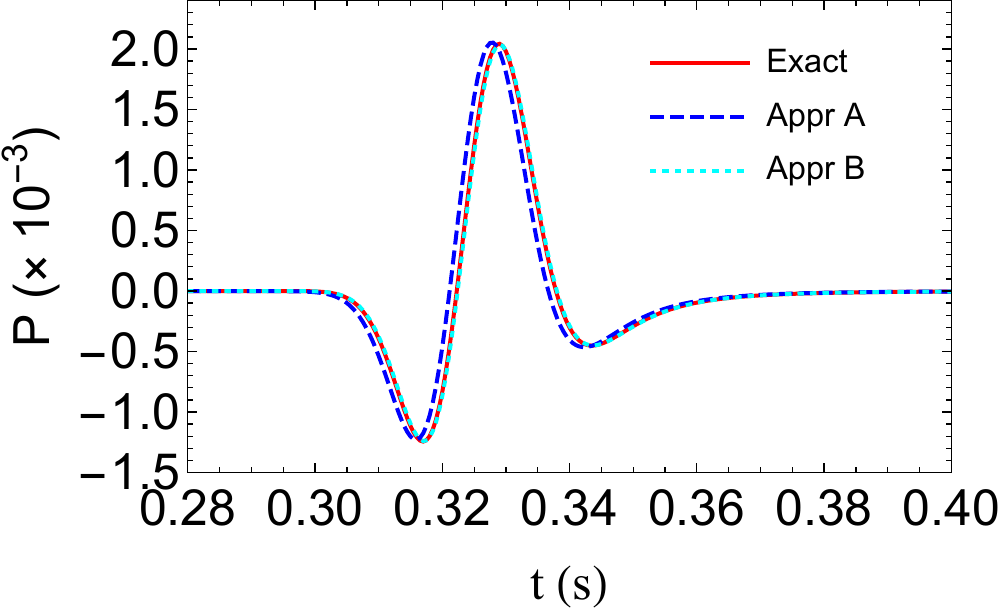}} 
\caption{
Comparison between the P-waveforms from a point source at $R=1$~km computed from the exact solution (``Exact'' in the legend) and approximate solutions A (``Appr A'' in the legend) and B (``Appr B'' in the legend). All three solutions are described in the main text. The group angle is: (a) $\psi=0\degree$, (b)  $\psi=30\degree$, (c) $\psi=60\degree$, and (d) $\psi=90\degree$. The medium parameters at $f_{0}=40$~Hz are $V_{P0}=3$~km/s, $\epsilon=0.1$, $\mathcal{A}_{P0}=0.025$ (corresponding to $Q_{33}=20$), and $\epsilon_{Q}=-0.3$. 
}
\label{fig:wave1}
\end{figure}

The source signal (whose spectrum is part of the wave equation \ref{eq:weq_freq}) is a 40~Hz Ricker wavelet. We compare the P-waveforms computed from the exact point-source solution (equations \ref{eq:P} and \ref{eq:tau}) and from approximate solutions that we call ``A'' (equation \ref{eq:Pnew} and equations \ref{eq:Vp_ellip}--\ref{eq:beta}) and ``B'' (similar to approximation A, but we use the exact group-velocity equation \ref{eq:Vp_ellip_exact} instead of approximation \ref{eq:Vp_ellip}). Figure \ref{fig:wave1} shows that the waveforms generated by the approximate and exact solutions practically coincide.

\begin{figure}[H]
\centering
\subfloat[]{\includegraphics[width=2.1in]{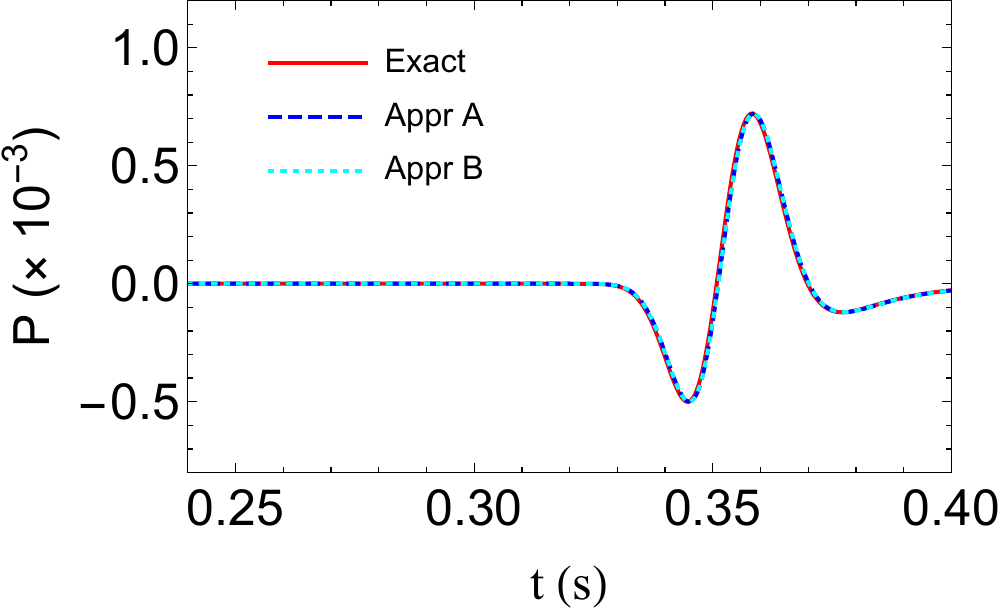}}  \qquad
\subfloat[]{\includegraphics[width=2.1in]{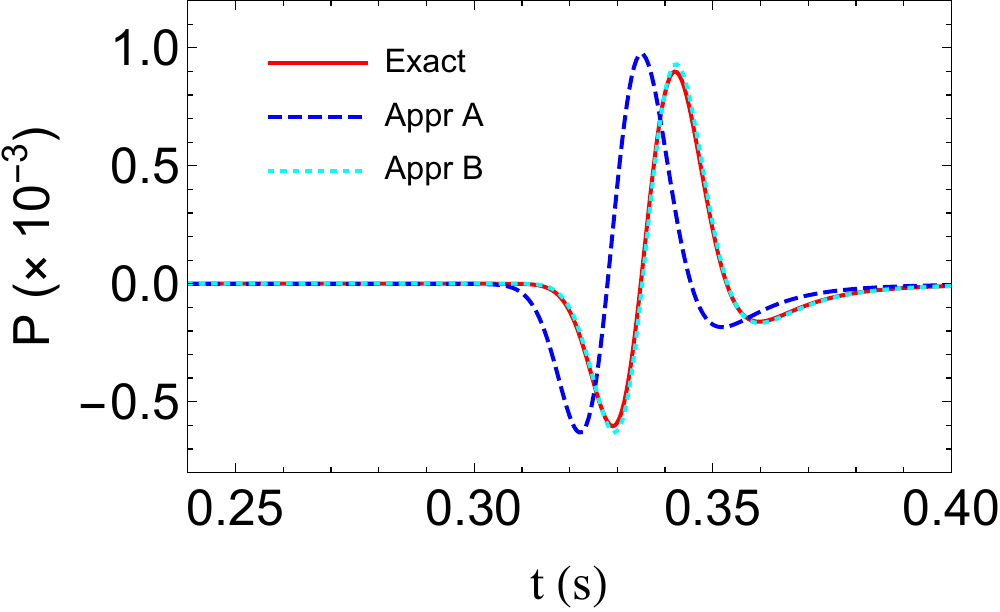}}  \\
\subfloat[]{\includegraphics[width=2.1in]{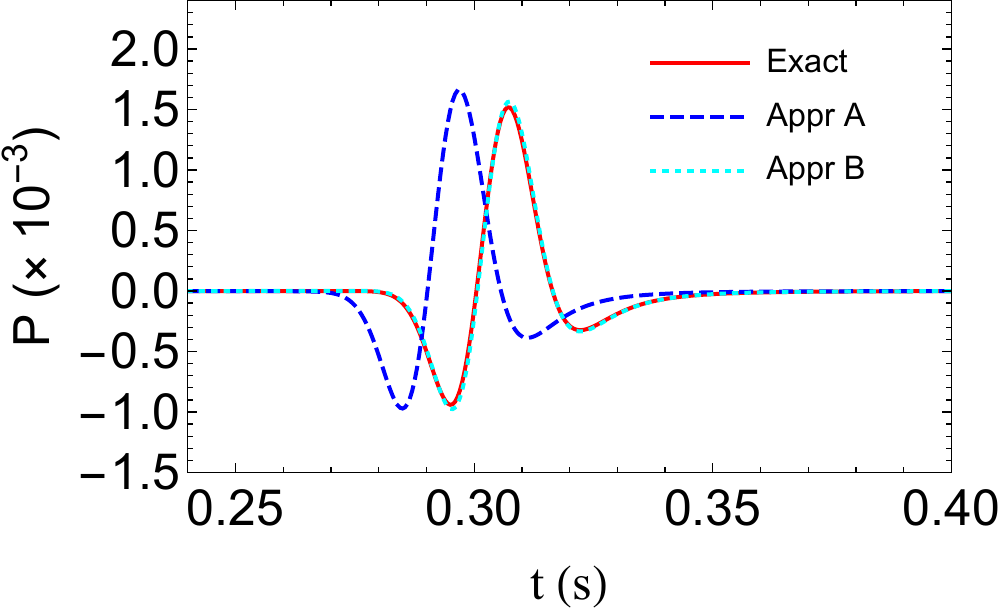}}  \qquad
\subfloat[]{\includegraphics[width=2.1in]{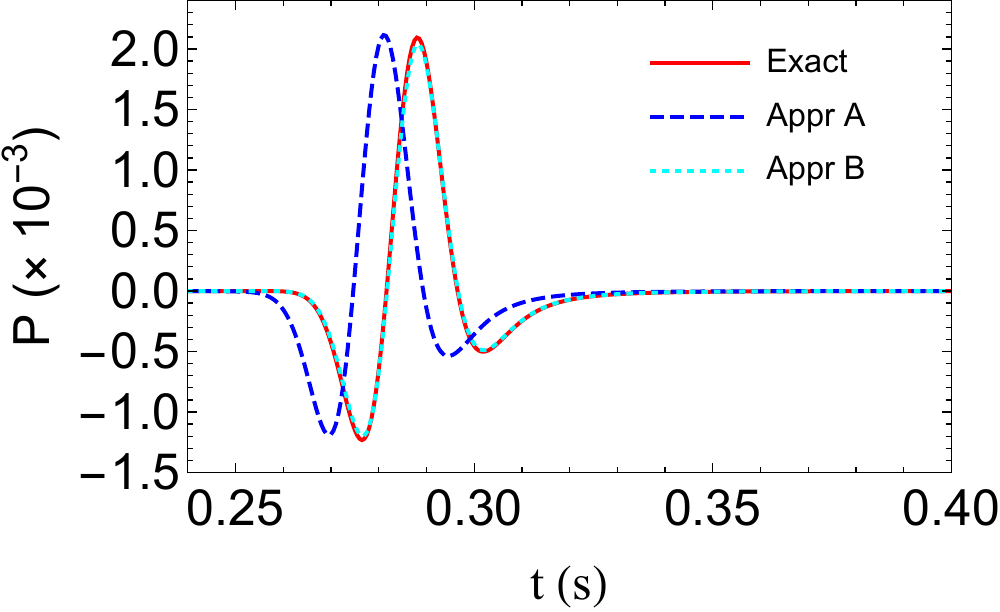}} 
\caption{
Same as Figure \ref{fig:wave1} but for $\epsilon=0.3$. 
}
\label{fig:wave2}
\end{figure}

Note that the late-arriving trough of the signal is attenuated more significantly than the earlier trough, which is consistent with the published results for viscoacoustic \cite[]{hao.greenhalgh:2021a} and  viscoelastic models \cite[]{bai.tsvankin:2016,hao.greenhalgh:2021b,hao.greenhalgh:2022b}.   
Figure \ref{fig:wave2} shows that approximate solution A becomes less accurate with increasing parameter $\epsilon$ but solution B is still close to the exact waveform. This result means that the error of the approximate solution is largely caused by using a linearized approximation for the group velocity. 

\begin{figure}[H]
\centering
\subfloat[]{\includegraphics[width=2.1in]{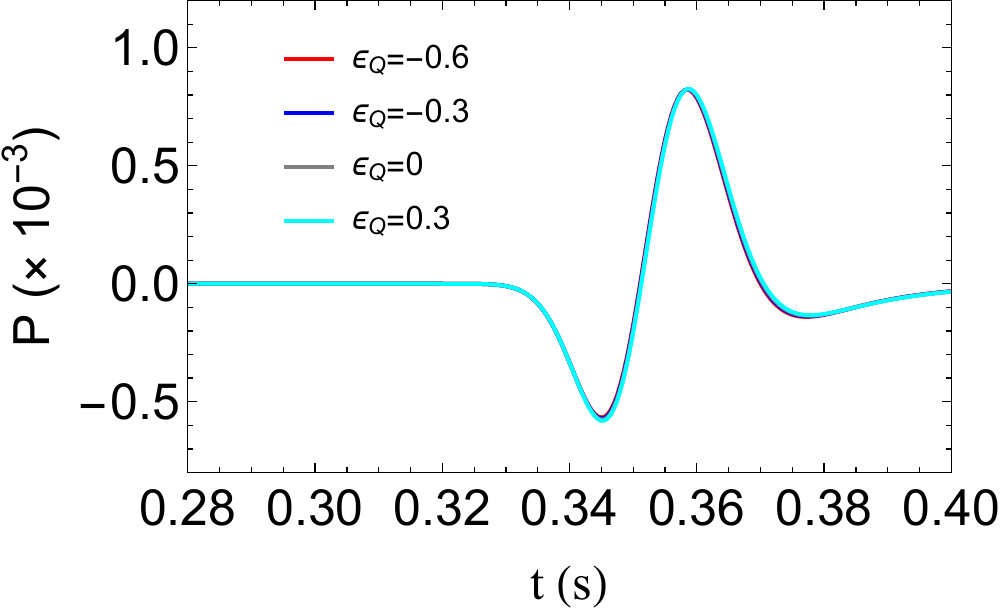}}  \qquad
\subfloat[]{\includegraphics[width=2.1in]{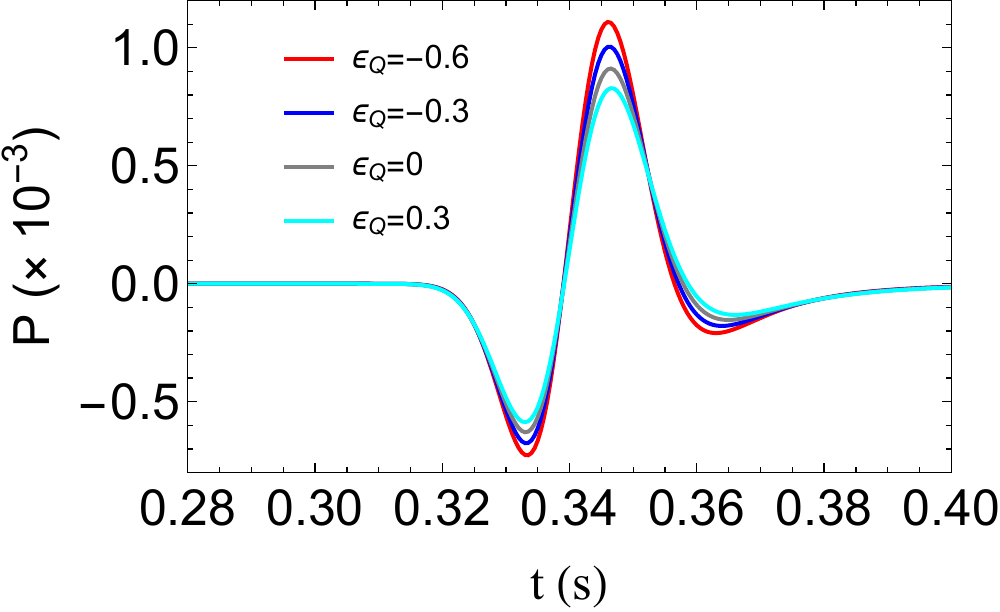}}  \\
\subfloat[]{\includegraphics[width=2.1in]{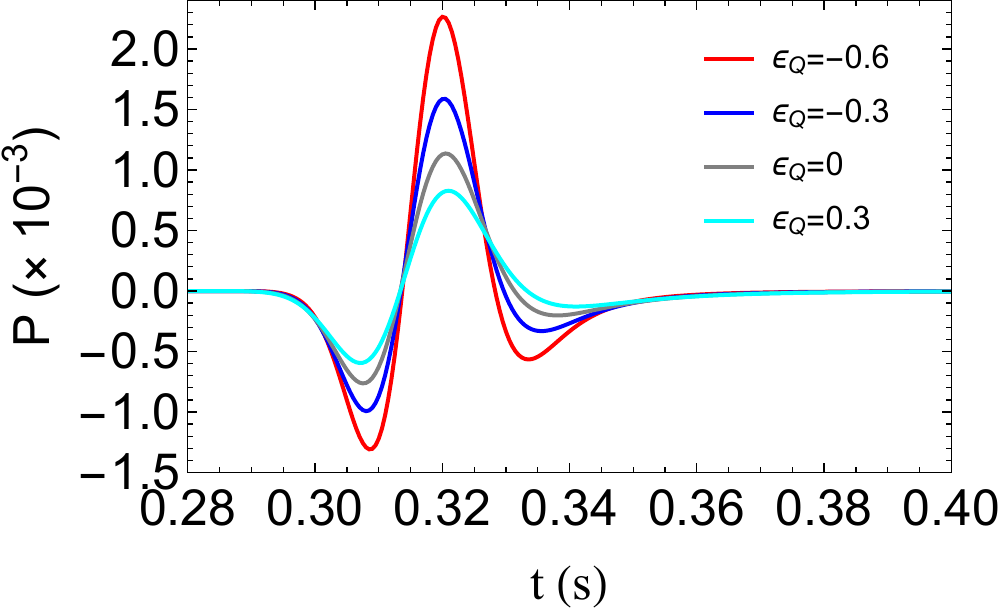}}  \qquad
\subfloat[]{\includegraphics[width=2.1in]{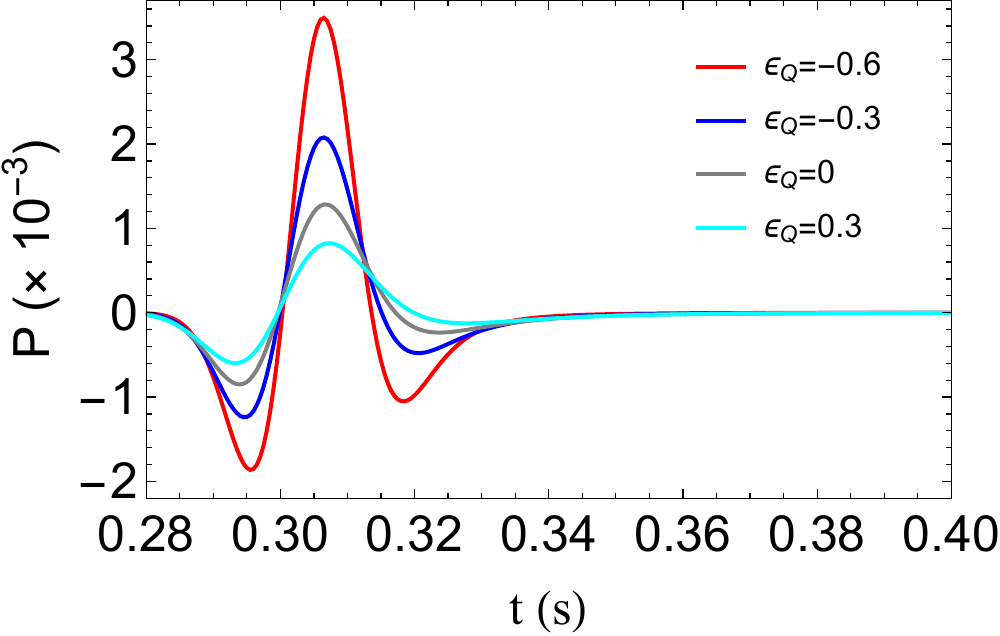}} 
\caption{
Point-source waveforms in constant-$Q$ elliptically anisotropic VTI media for different values of the parameter $\epsilon_{Q}$. The other medium parameters at $f_{0}=40$~Hz are $V_{P0}=3$~km/s, $\epsilon=0.2$, and 
$\mathcal{A}_{P0}=0.025$ (corresponding to $Q_{33}=20$). 
The group angle is (a) $\psi=0\degree$, (b)  $\psi=30\degree$, (c) $\psi=60\degree$, and (d) $\psi=90\degree$. 
}
\label{fig:wave_epsQ}
\end{figure}

Finally, we use the exact point-source solution to analyze the influence of the attenuation-anisotropy parameter $\epsilon_{Q}$ for different propagation directions. Figure \ref{fig:wave_epsQ} confirms that the coefficient $\epsilon_{Q}$ influences the amplitude and waveform of P-waves in oblique propagation directions, especially close to the isotropy plane that corresponds to $\psi=90\degree$. In particular, large negative values of $\epsilon_{Q}$ substantially reduce attenuation in near-horizontal directions (see equation \ref{eq:Ap2}). However, in agreement with previous publications, $\epsilon_{Q}$ has practically no impact on P-wave velocity \cite[]{zhu.tsvankin:2006,HaoA2017a}.  

\section{Conclusions}
We obtained two elliptical conditions for attenuative transversely isotropic media expressed in terms of the Thomsen velocity parameters and Thomsen-type attenuation parameters. The first condition, which is the same as in purely elastic media ($\epsilon=\delta$), makes the P-wave slowness surface elliptical. The second condition ensures that the inverse of the P-wave normalized phase attenuation coefficient approximately satisfies an elliptical equation. These two conditions lead to the elliptical form of the P-wave dispersion relation expressed through the complex wavenumbers. 

Also, we employed the correspondence principle to derive the point-source solution of the viscoacoustic wave equation for elliptical VTI media. Assuming weak velocity and attenuation anisotropy, the obtained solution was approximately expressed through the geometric-spreading factor, group or phase velocity, phase attenuation coefficient, and the attenuation-related phase shift. The accuracy of this approximation for moderate values of the anisotropy parameters can be increased by using the exact group-velocity function.  
The analytic results presented here provide the foundation for using elliptical anisotropy as the reference model in perturbation-based solutions for general attenuative TI media.

%

\begin{appendix}
\section{Appendix A: Point-source solution of viscoacoustic wave equation}
Here, we derive the point-source solution of the viscoacoustic wave equation \ref{eq:weq} for elliptically anisotropic VTI media. The frequency-domain acoustic wave equation for nonattenuative elliptical anisotropy is given by:
\begin{equation} \label{eq:wef}
-\omega^2 \hat{P}' = a_{11}  \left( \frac{\partial^2 \hat{P}'}{\partial x^2} + \frac{\partial^2 \hat{P}'}{\partial y^2} \right) +  a_{33} \frac{\partial^2 \hat{P}'}{\partial z^2}  + \hat{S}(\omega) \delta(x) \delta(y) \delta(z)  ,
\end{equation}
where $\hat{P}'$ is the frequency-domain wavefield, and $a_{11}$ and $a_{33}$ are the real-valued stiffness coefficients $c_{11}$ and $c_{33}$ normalized by density. 

Using the coordinate transformation
\begin{equation}
\tilde{x} =  \frac{x}{ \sqrt{a_{11}}} ,  \quad
\tilde{y} =  \frac{y}{\sqrt{a_{11}}} ,  \quad
\tilde{z} =  \frac{z}{\sqrt{a_{11}}} ,  
\end{equation}
equation \ref{eq:wef} can be rewritten as:
\begin{equation} \label{eq:awe}
-\omega^2 \hat{P}' = \tilde{\nabla}^2 \hat{P}' + \frac{\hat{S}(\omega)}{ a_{11} \sqrt{a_{33} } } \delta(\tilde{x}) \delta(\tilde{y}) \delta(\tilde{z})  , 
\end{equation}
where $ \tilde{\nabla}^2$ denotes the Laplacian operator:
\begin{equation} \label{eq:xx}
\tilde{\nabla}^2 = \frac{\partial^2 }{\partial \tilde{x}^2}  + \frac{\partial^2 }{\partial \tilde{y}^2} + \frac{\partial^2 }{\partial \tilde{z}^2} .
\end{equation}
Equation \ref{eq:awe} takes advantage of the scaling property of the Dirac delta function \cite[]{arfken:2013}. 

The  solution of equation \ref{eq:awe}, which is equivalent to the conventional acoustic wave equation, is \cite[]{aki.richards:2002}:
\begin{equation} \label{eq:Pprime}
\hat{P}' = \frac{\hat{S}(\omega)}{4\pi  a_{11} \sqrt{a_{33}} \tau'} 
\exp(-i\omega \tau' ) ,
\end{equation}
where 
\begin{equation}
\tau' = \sqrt{ \frac{x^2}{a_{11}} + \frac{y^2}{a_{11}} + \frac{z^2}{a_{33}} } \, . 
\end{equation}

The frequency-domain wave equation for attenuative elliptically anisotropic media has the form (see equation \ref{eq:weq_freq}): 
\begin{equation} \label{eq:weq_freq2}
-\omega^2 \hat{P} = m_{11}  \left( \frac{\partial^2 \hat{P}}{\partial x^2} + \frac{\partial^2 \hat{P}}{\partial y^2} \right) +  m_{33} \, \frac{\partial^2 \hat{P}}{\partial z^2}  + \hat{S}(\omega) \delta(x) \delta(y) \delta(z)  .  
\end{equation}
According to the correspondence principle \cite[]{ben-menahem.singh:1981,carcione:2014}, an analytic solution of any time-harmonic problem for attenuative media has the same form as that of the corresponding problem for the reference purely elastic medium. Hence, by analogy with equation \ref{eq:Pprime}, the point-source solution of the viscoacoustic wave equation \ref{eq:weq_freq2} can be found as:
\begin{equation}
\hat{P} = \frac{\hat{S}(\omega)}{4\pi  m_{11} \sqrt{m_{33}} \tau} 
\exp(-i\omega \tau ) , 
\end{equation}
where
\begin{equation}
\tau = \sqrt{ \frac{x^2}{m_{11}} + \frac{y^2}{m_{11}} + \frac{z^2}{m_{33}} } \, . 
\end{equation}

\end{appendix}

\bibliographystyle{macros/elsarticle-num}
\bibliography{refs/refs20221221,refs/qi_refs20220719}

\begin{thebibliography}{10}
\expandafter\ifx\csname url\endcsname\relax
  \def\url#1{\texttt{#1}}\fi
\expandafter\ifx\csname urlprefix\endcsname\relax\def\urlprefix{URL }\fi
\expandafter\ifx\csname href\endcsname\relax
  \def\href#1#2{#2} \def\path#1{#1}\fi

\bibitem{thomsen:1986}
L.~Thomsen, Weak elastic anisotropy, Geophysics 51~(10) (1986) 1954--1996.

\bibitem{tsvankin:2012}
I.~Tsvankin, Seismic signatures and analysis of reflection data in anisotropic
  media (3rd ed.), Society of Exploration Geophysicists, 2012.

\bibitem{slawinski:2004}
M.~A. Slawinski, C.~J. Wheaton, M.~Powojowski, Vsp traveltime inversion for
  linear inhomogeneity and elliptical anisotropy, Geophysics 69~(2) (2004)
  373--377.

\bibitem{danek:2012}
T.~Danek, M.~A. Slawinski, Bayesian inversion of vsp traveltimes for linear
  inhomogeneity and elliptical anisotropy, Geophysics 77~(6) (2012) R239--R243.

\bibitem{stovas:2012}
A.~Stovas, T.~Alkhalifah, A new traveltime approximation for {TI} media,
  Geophysics 77~(4) (2012) C37--C42.

\bibitem{helbig:1983}
K.~Helbig, Elliptical anisotropy—its significance and meaning, Geophysics
  48~(7) (1983) 825--832.

\bibitem{rogister:2005}
Y.~Rogister, M.~A. Slawinski, Analytic solution of ray-tracing equations for a
  linearly inhomogeneous and elliptically anisotropic velocity model,
  Geophysics 70~(5) (2005) D37--D41.

\bibitem{levin:1978}
F.~K. Levin, The reflection, refraction, and diffraction of waves in media with
  an elliptical velocity dependence, Geophysics 43~(3) (1978) 528--537.

\bibitem{daley:1979}
P.~Daley, F.~Hron, Reflection and transmission coefficients for seismic waves
  in ellipsoidally anisotropic media, Geophysics 44~(1) (1979) 27--38.

\bibitem{verwest:1989}
B.~J. Ver{W}est, Seismic migration in elliptically anisotropic media,
  Geophysical prospecting 37~(2) (1989) 149--166.

\bibitem{schleicher:2008}
J.~Schleicher, A.~Novais, J.~Costa, Vertical image waves in elliptically
  anisotropic media, Studia Geophysica et Geodaetica 52~(1) (2008) 101--122.

\bibitem{zhu.tsvankin:2007b}
Y.~Zhu, I.~Tsvankin, P.~Dewangan, K.~van Wijk, Physical modeling and analysis
  of p-wave attenuation anisotropy in transversely isotropic media, Geophysics
  72~(1) (2007) D1--D7.

\bibitem{zhu:2007}
Y.~Zhu, I.~Tsvankin, I.~Vasconcelos, Effective attenuation anisotropy of
  thin-layered media, Geophysics 72~(5) (2007) D93--D106.

\bibitem{shekar:2011}
B.~Shekar, I.~Tsvankin, Estimation of shear-wave interval attenuation from
  mode-converted data, Geophysics 76~(6) (2011) D11--D19.

\bibitem{shekar:2012}
B.~Shekar, I.~Tsvankin, Anisotropic attenuation analysis of crosshole data
  generated during hydraulic fracturing, The Leading Edge 31~(5) (2012)
  588--593.

\bibitem{vavryvcuk:2007asympt}
V.~Vavry\v{c}uk, Asymptotic {G}reen's function in homogeneous anisotropic
  viscoelastic media, Proceedings of the Royal Society A: Mathematical,
  Physical and Engineering Sciences 463~(2086) (2007) 2689--2707.

\bibitem{shekar.tsvankin:2014}
B.~Shekar, I.~Tsvankin, Point-source radiation in attenuative anisotropic
  media, Geophysics 79~(5) (2014) WB25--WB34.

\bibitem{hao.alkhalifah:2019}
Q.~Hao, T.~Alkhalifah, Viscoacoustic anisotropic wave equations, Geophysics
  84~(6) (2019) C323--C337.

\bibitem{vavrycuk:2007}
V.~Vavry\v{c}uk, Ray velocity and ray attenuation in homogeneous anisotropic
  viscoelastic media, Geophysics 72~(6) (2007) D119--D127.

\bibitem{behura:2009b}
J.~Behura, I.~Tsvankin, Role of the inhomogeneity angle in anisotropic
  attenuation analysis, Geophysics 74~(5) (2009) WB177--WB191.

\bibitem{zhu.tsvankin:2006}
Y.~Zhu, I.~Tsvankin, Plane-wave propagation in attenuative transversely
  isotropic media, Geophysics 71~(2) (2006) T17--T30.

\bibitem{zhu.tsvankin:2007}
Y.~Zhu, I.~Tsvankin, Plane-wave attenuation anisotropy in orthorhombic media,
  Geophysics 72~(1) (2007) D9--D19.

\bibitem{hao.tsvankin:2022}
Q.~Hao, I.~Tsvankin, Thomsen-type parameters and attenuation coefficients for
  constant-{$Q$} transverse isotropy, Geophysics. Submitted for Peer Review.

\bibitem{cerveny:2001}
V.~{\v{C}}erven{\'y}, Seismic ray theory, Cambridge University Press, 2001.

\bibitem{tsvankin.grechka:2011}
I.~Tsvankin, V.~Grechka, Seismology of azimuthally anisotropic media and
  seismic fracture characterization, Society of Exploration Geophysicists,
  2011.

\bibitem{arfken:2013}
G.~Arfken, H.~Weber, F.~Harris, Mathematical Methods for Physicists: A
  Comprehensive Guide, Elsevier Science, 2013.

\bibitem{hudson:1980}
J.~A. Hudson, The excitation and propagation of elastic waves, Cambridge
  University Press, 1980.

\bibitem{hao.greenhalgh:2019}
Q.~Hao, S.~Greenhalgh, The generalized standard-linear-solid model and the
  corresponding viscoacoustic wave equations revisited, Geophysical Journal
  International 219~(3) (2019) 1939--1947.

\bibitem{hao.greenhalgh:2021b}
Q.~Hao, S.~Greenhalgh, Nearly constant {$Q$} dissipative models and wave
  equations for general viscoelastic anisotropy, Proceedings of the Royal
  Society A: Mathematical, Physical and Engineering Sciences 477~(2251) (2021)
  20210170.

\bibitem{hao.greenhalgh:2021a}
Q.~Hao, S.~Greenhalgh, Nearly constant {$Q$} models of the generalized standard
  linear solid type and the corresponding wave equations, Geophysics 86~(4)
  (2021) T239--T260.

\bibitem{ben-menahem.singh:1981}
A.~Ben-Menahem, S.~J. Singh, Seismic waves and sources, Springer-Verlag, 1981.

\bibitem{carcione:2014}
J.~M. Carcione, Wave fields in real media: Theory and numerical simulation of
  wave propagation in anisotropic, anelastic, porous and electromagnetic media:
  Handbook of Geophysical exploration (3rd ed.), Elsevier, 2014.

\bibitem{aki.richards:2002}
K.~Aki, P.~Richards, Quantitative {S}eismology (2nd ed.), University Science
  Books, 2002.

\bibitem{stovas:2018}
A.~Stovas, Geometric spreading in orthorhombic media, Geophysics 83~(1) (2018)
  C61--C73.

\bibitem{tsvankin:1995}
I.~Tsvankin, Body-wave radiation patterns and {AVO} in transversely isotropic
  media, Geophysics 60~(5) (1995) 1409--1425.

\bibitem{kjartansson:1979}
Kjartansson, Constant {$Q$}-wave propagation and attenuation, Journal of
  Geophysical Research 84 (1979) 4737--4748.

\bibitem{bai.tsvankin:2016}
T.~Bai, I.~Tsvankin, Time-domain finite-difference modeling for attenuative
  anisotropic media, Geophysics 81~(2) (2016) C69--C77.

\bibitem{hao.greenhalgh:2022b}
Q.~Hao, S.~Greenhalgh, X.~Huang, H.~Li, Viscoelastic wave propagation for
  nearly constant {$Q$} transverse isotropy, Geophysical Prospecting 70~(7)
  (2022) 1176--1192.

\bibitem{HaoA2017a}
Q.~Hao, T.~Alkhalifah, An acoustic eikonal equation for attenuating
  transversely isotropic media with a vertical symmetry axis, Geophysics 82~(1)
  (2017) C9--C20.

\end{thebibliography}

\end{document}